# Design and Construction of a First Prototype Muon Tomography System with GEM Detectors for the Detection of Nuclear Contraband

M. Hohlmann, *Member, IEEE*, K. Gnanvo, *Member, IEEE*, L. Grasso, J. B. Locke, A. Quintero, *Member, IEEE* and D. Mitra, *Senior Member, IEEE*

*Abstract–* Current radiation portal monitors at sea ports and international borders that employ standard radiation detection techniques are not very sensitive to nuclear contraband that is well shielded to absorb emanating radiation. Muon Tomography (MT) based on the measurement of multiple scattering of atmospheric cosmic ray muons traversing cargo or vehicles that contain high-Z material is a promising passive interrogation technique for solving this problem. We report on the design and construction of compact Micro-Pattern Gas Detectors for a small prototype MT station. This station will employ 10 tracking stations based on 30cm × 30cm low-mass triple-GEM detectors with 2D readout. Due to the excellent spatial resolution of GEMs it is sufficient to use a gap of only a few cm between tracking stations. Together with the compact size of the GEM detectors this allows the GEM MT station to be an order of magnitude more compact than MT stations using traditional drift tubes. We present details of the production and assembly of the GEM-based tracking stations in collaboration with CERN and the RD51 collaboration as well as the design of the initial front-end electronics and readout system.

## I. Introduction

Muon Tomography (MT) based on the measurement of multiple scattering of atmospheric cosmic ray muons has been suggested as a promising technique for detecting shielded high-Z material in cargo and discriminating it from low-Z background material [1]-[6]. Muons are charged elementary particles with mass 105.7 MeV/$c^2$ and get produced in the upper atmosphere by primary cosmic rays. The muon flux at sea level is $\sim 10^4$ min$^{-1}$ m$^{-2}$ at an average energy of 4 GeV. Multiple Coulomb scattering depends on density and atomic number $Z$ of the material traversed. Due to their penetrating nature, cosmic ray muons are good candidates for detecting shielded high-Z materials. Nuclear materials that pose a homeland security threat typically have high atomic numbers ($Z > 82$) so that multiple scattering of muons can be exploited for a non-invasive scanning system to detect such materials. Computer simulations have shown that cosmic ray muon tomography can discriminate sensitive high-Z nuclear material such as uranium against iron or steel background with high statistical significance in 4-10 minutes of exposure if the detectors have good spatial resolution of $\sim$100 µm [7], [8].

The Gas Electron Multiplier (GEM) detector is a micro-pattern gas detector for charged particles [9]. It uses a thin foil of Kapton coated with copper layers on both sides and pierced by a regular array of chemically etched holes, typically 140 µm apart. A voltage is applied across such a GEM foil and the resulting high electric field in the holes can make an avalanche of ions and electrons pour through the holes. The electrons are collected by a suitable device; here a readout plane with x-y strips.

We are currently constructing a small prototype of a muon tomography station with GEM detectors to experimentally determine its performance.

## II. GEM Detector Assembly

We are using 30cm × 30cm GEM foils based on an upgraded version of the original foils used by the COMPASS experiment at CERN [10], but with the central foil area also sensitive to traversing particles. This updated design was originally proposed by the TERA Foundation for a medical application [11]. The GEM foils have 12 sectors, so in case of a short one loses only one sector instead of an entire foil.

All components used for the GEM detector construction material and the readout electronics have been selected to minimize the mass and consequently multiple scattering in the GEM detectors themselves. Each detector is fabricated by gluing together a set of thin frames holding the various foils and electrodes in a similar procedure as used for assembly of the COMPASS [10] and TERA [11] GEM detectors, but with some small improvements of the assembly techniques. Mechanical stiffness and gas tightness are provided by two low-mass outer honeycomb supporting plates

### A. High voltage test of GEM foils

The acceptance criterion for a GEM foil requires the foil to hold 500 V under nitrogen gas with a leakage current less than 5nA in each of the 12 HV sectors. These tests are made

Manuscript received November 13, 2009. This work was supported in part by the U.S. Department of Homeland Security under Grant No. 2007-DN-077-ER0006-02.

M. Hohlmann, K. Gnanvo, L. Grasso, J. B. Locke, and A. Quintero are with the Department of Physics and Space Sciences, Florida Institute of Technology, Melbourne, FL 32901, USA (telephone: 321-674-7275, e-mail: hohlmann@fit.edu).

D. Mitra is with the Department of Computer Science, Florida Institute of Technology, Melbourne, FL 32901, USA.

in a class 1000 clean room and are performed before and after framing the foils (figure 1). The tests are performed in an air-tight Plexiglas box with nitrogen flowing at a rate of 5 volume exchanges in 3 hours; a CAEN N471A high voltage power supply and a Tabor Electronics DMM4030 ammeter are used for the current measurements. A total of 30 foils were delivered by the CERN PCB workshops and all of them passed the HV test before framing, with an average leakage current of 2.5 nA. Out of 24 framed foils, 18 passed the HV test with an average of 1.3 nA; five foils showed unacceptably large currents and are currently under investigation; one foil was lost due to stretching problems.

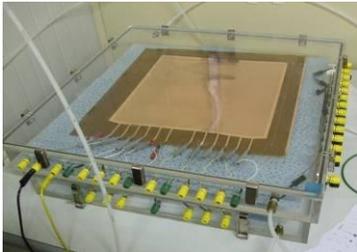

Fig. 1. GEM foil under HV test in an air-tight Plexiglas box under nitrogen at the GDD lab at CERN.

### B. Frames for the GEM foils

The frames for the foils provide the mechanical tension to the foils and their thickness is used to separate the foils to obtain the required electric fields. As in the standard CERN triple-GEM detector design, we are using a 3mm spacer for the Drift Cathode to establish the distance between drift cathode and first GEM and 2mm spacers between the GEM foils and between third GEM foil and readout board.

These frames come from the CERN PCB machine shop with two areas of extra material; the outer one is the periphery of the FR4 plate used to make the frames and is removed before working with the frames (figure 2). The middle area allows for alignment pins and is designed to be relatively wide so that strong adhesion of the foil to the frame and good foil flatness during the stretching process can be guaranteed. This results in an improvement in the GEM framing process. The grid structure of the frame is sanded down and coated with 2-component polyurethane Nuvovern LW of Walter Mader AG to prevent dust particles from the frame itself to get on the GEM foil.

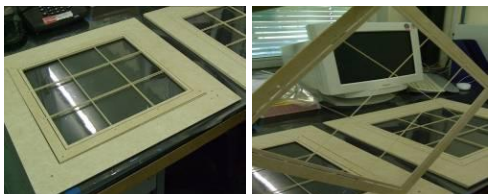

Fig. 2. Raw frame delivered by the CERN machine shop (left), frames used for stretching the GEM foils without the outer material part (right).

### C. Foil stretching process

We use a thermal method for tensioning the GEM foils. The foils are sandwiched between two Plexiglas frames and put into an oven at 45°C, which stretches the foil. We use ARALDIT AY103+HD991 (ratio 10:4) from Cyba-Geigi to glue a frame onto the tensioned foil and then put the assembly again into the oven to maintain tension for 12 hours until the glue is fully cured.

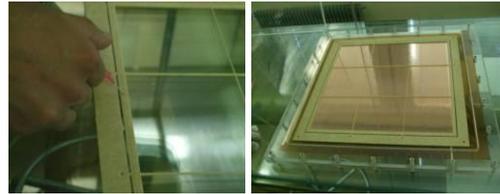

Fig. 3. Frames for the foils (left), framed foil in the thermal stretching device ready to go into the oven to cure the glue (right).

### D. Drift Cathode

The mechanical rigidity of the detectors is provided by two honeycomb plates, one supporting the readout plane and the other supporting the drift cathode. To glue the drift onto the honeycomb plate, we sparingly apply glue uniformly to the kapton side of the drift cathode and to the honeycomb plate and then press them together. Air bubbles between the honeycomb and the drift must be removed using a roller before putting heavy weights onto the assembly. After 12 hours, the glue is cured and the spacer between the drift and the first GEM foil is glued onto the drift cathode using the alignment pins in the outer frame. Care must be taken that the corner of the gas inlet is on top of the drift high voltage strip.

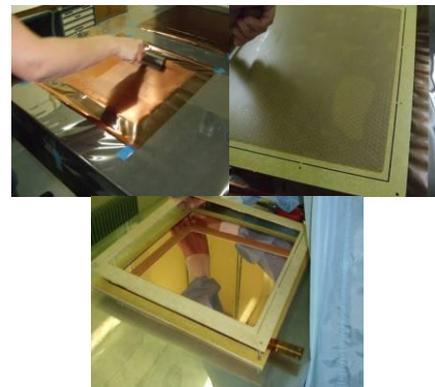

Fig. 4. Gluing the Kapton side of the drift (left); undesirable air bubble after gluing the drift cathode and honeycomb plate (right); gluing the drift cathode spacer (bottom).

### E. Readout

The gluing of the readout foil works the same way as for the drift cathode, but the honeycomb structure of the readout has the gas outlet of the chamber and must be glued in a specific way (figure 5). After gluing, the readout goes to the machine shop so that its final shape can be cut.

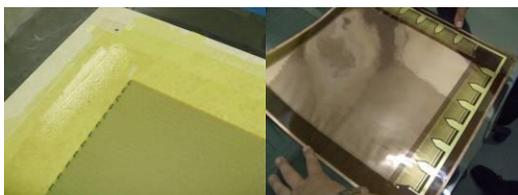

Fig. 5. Honeycomb structure of the readout with the gas outlet (left), readout with 2-D strips (right).

*F. Final assembly of the complete detector*

Using the alignment pins on the drift, the spacer of the drift is glued to the first framed GEM foil. Glue is applied to the frame of the first GEM and the second GEM foil is placed on top. This is repeated for the third GEM and a heavy weight is put on the entire assembly while the glue cures (figure 6). Finally, we remove the alignment frame from the GEM stack and glue the stack onto the readout plane.

Gas connectors are glued into the frames and the edges of the detector assembly are sealed with 1-2577 conformal coating (Dow Corning) to prevent gas leaks. The detector is taken to the machine shop to solder twelve 130-pin Panasonic AXK6SA3677YG connectors onto the periphery of the readout plane to provide connectivity to the x-y strips.

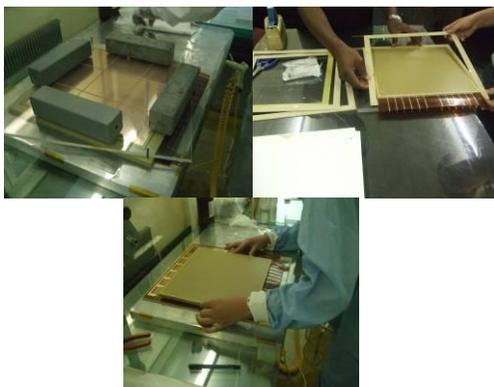

Fig. 6. GEM stack glued with weight (left); taking out the alignment frames (right); gluing the GEM stack onto the readout plane (bottom).

*G. High Voltage circuit board*

The design of the high voltage (HV) circuit is basically a simple voltage divider. Since the GEM foils have 12 separate sectors, the high voltage circuit has 12 separate sections for each foil (figure 7 top).

We are using a HV board designed for the TERA Foundation GEM detectors, which itself is an upgrade of the COMPASS HV board. The circuit board is manufactured by the CERN PCB machine shop and the resistors are manually soldered onto it (figure 7 bottom). Before mounting it to the detector, the boards are tested by taking the main supply voltage up to 5 kV and measuring the bias current to verify that the boards have proper Ohmic behavior. The assembled boards are cleaned, coated, and retested.

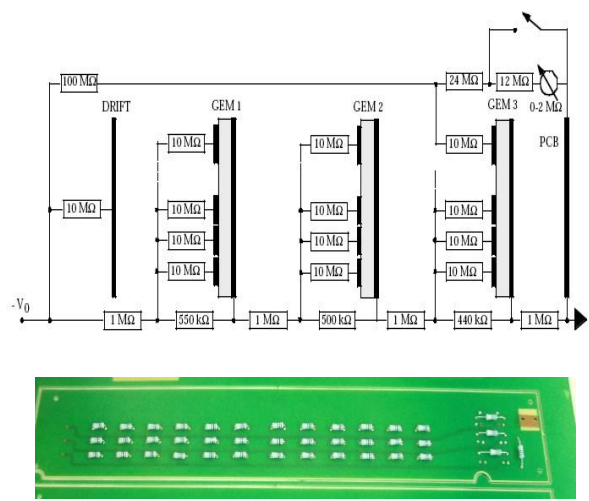

Fig. 7. High voltage circuit diagram [10] (top); printed circuit board with resistors soldered in (bottom).

We assembled 6 HV boards in this way and all of them passed the tests without problems. The last step of the assembly is to attach the HV board to the honeycomb base plate and to solder the strip electrodes of the GEM foils to the HV board.

### III. INITIAL READOUT ELECTRONICS

The analog front-end amplifier is based on the 96-channel "Gassiplex" chip, which was developed for the CAST experiment at CERN [12]. We have developed an adapter card to make the interface between the Gassiplex front-end and our detectors.

We are using a VME-based Data Acquisition (DAQ) with 8 CAEN CRAMs and a data sequencer. The CRAM modules receive the data signal from the Gassiplex cards (two Gassiplex per CRAM). The sequencer card receives the trigger signal, produces the control signals for the Gassiplex and for the CRAMs, receives a Data Ready signal if there are data available on the CRAMs, and clears the CRAMs modules at the end of an event readout. The sequencer card is connected to a computer and the acquired signal is read out with LabView software. We are currently upgrading the DAQ in order to accommodate up to 16 Gassiplex cards because the original CAST software [13] cannot read out more than 4 ADCs.

### IV. FIRST PROTOTYPE FOR A MUON TOMOGRAPHY STATION

A simple design was chosen for a mechanical stand for our first prototype station that will accommodate multiple top and bottom GEM detectors with 30cm × 30cm active areas (figure 8). The stand can be adjusted to study the effect that various detector gaps have on the tomographic imaging. We are planning to design a second stand that can accommodate GEM detectors also on two vertical sides giving an imaging volume with detectors on a total of four sides.

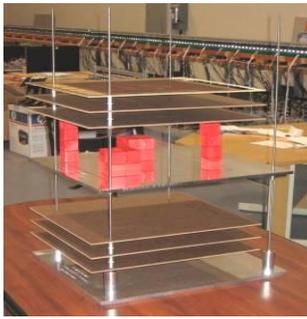

Fig. 8. Mechanical stand for first small MT prototype station with GEM detector and target mock-ups (red).

## V. GEM Detector Commissioning

The detectors were shielded against electric noise before testing. They were initially tested under HV at 100% $CO_2$ and then operated with an $Ar/CO_2$ 70:30 counting gas mixture. The detectors were placed on an X-ray test bench and at a total bias high voltage of 3.8 kV signal pulses become visible. A typical pulse height spectrum obtained with the GEM detectors exposed to 8 keV X-rays is shown in figure 9. A total of three detectors were tested with this procedure and all of them show similar behavior. Not a single spark was observed during any of the tests and the signal is acquired with low electrical noise for all three assembled detectors.

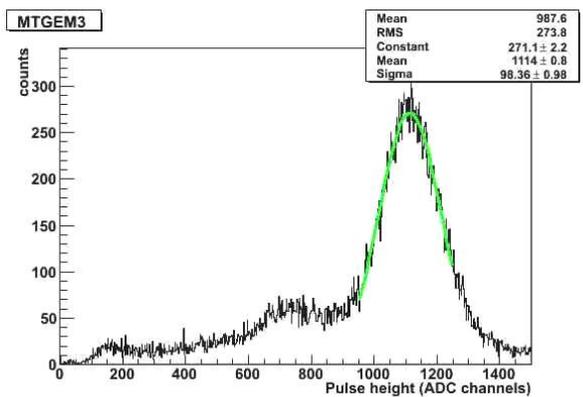

Fig. 9. Energy spectrum obtained showing a ~ 20% energy resolution (FWHM) for 8 keV X-ray.

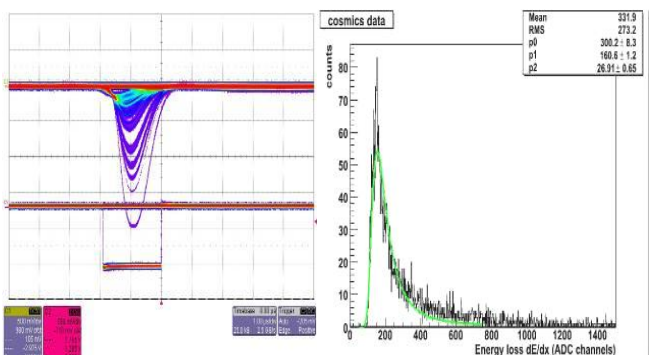

Fig. 10. Minimum ionizing pulses using cosmic ray muons recorded with one GEM detector in single channel mode (left). Corresponding pulse height distribution with fit to Landau curve in green (right).

Cosmic ray muon data were also collected with one of the detectors. 100,000 muon events were recorded using 1/6 of the total active area (with all 128 strips from one Panasonic connector ganged together in the readout) for 5 hours. Raw pulses and the pulse height spectrum are shown in figure 10.

## VI. Conclusions and Outlook

Muon tomography based on multiple Coulomb scattering of cosmic ray muons appears as a promising way to distinguish high-Z threat materials in cargo from low-Z and medium-Z background with high statistical significance. We are currently building a first MT station prototype with 30cm × 30cm GEM detectors to demonstrate the validity of using MPGDs in the tracking station for muon tomography. A total of six GEM detectors were assembled so far at the GDD lab at CERN. Preliminary tests of three detectors with X-rays show expected performance and similar behavior. Tests with cosmic ray muons conducted with one detector show satisfactory results with pulse heights following a Landau distribution as expected. We plan to get the first data from a GEM-based MT prototype station by early 2010.

The next major step in this project will be to build an MT station prototype based on large-area GEMs and to test it under realistic conditions for vehicle or container scanning. To do so we need larger GEM detectors (~ 100cm × 100cm) as the base unit for a tracking station. Efforts are being made by the RD51 collaboration for various HEP applications to build such large-area GEM detectors. We plan to participate in different aspects of the R&D for such large-area GEMs ranging from the framing and testing of the large GEM foils to the challenges associated with the electronic readout system needed for this detectors.

### Acknowledgment and Disclaimer

We thank Leszek Ropelewski and the GDD group, Rui de Oliveira and the PCB production facility, and Miranda Van Stenis, all at CERN, for their help and technical support with the detector construction. This material is based upon work supported in part by the U.S. Department of Homeland Security under Grant Award Number 2007-DN-077-ER0006-02. The views and conclusions contained in this document are those of the authors and should not be interpreted as necessarily representing the official policies, either expressed or implied, of the U.S. Department of Homeland Security.